\documentclass[aps,pra,groupedaddress,tightenlines]{revtex4}

\usepackage{amsmath}
\usepackage{graphicx}
\usepackage[english]{babel}

\begin{document}

\title{Continuous quantum error correction for non-Markovian decoherence}
\author{Ognyan Oreshkov}
\email{oreshkov@usc.edu}
\affiliation{Department of Physics, University of Southern California,
Los Angeles, CA  90089}
\author{Todd A. Brun}
\email{tbrun@usc.edu}
\affiliation{Communication Sciences Institute, University of Southern California,
Los Angeles, CA  90089}

\date{\today}

\begin{abstract}
We study the effect of continuous quantum error correction in the
case where each qubit in a codeword is subject to a general
Hamiltonian interaction with an independent bath. We first
consider the scheme in the case of a trivial single-qubit code,
which provides useful insights into the workings of continuous
error correction and the difference between Markovian and
non-Markovian decoherence. We then study the model of a bit-flip
code with each qubit coupled to an independent bath qubit
and subject to continuous correction, and find its solution. We show that for sufficiently large
error-correction rates, the encoded state approximately follows an
evolution of the type of a single decohering qubit, but with an
effectively decreased coupling constant. The factor
by which the coupling constant is decreased scales quadratically
with the error-correction rate. This is compared to the case of
Markovian noise, where the decoherence rate is effectively
decreased by a factor which scales only linearly with the rate of
error correction. The quadratic enhancement depends on the
existence of a Zeno regime in the Hamiltonian evolution which is
absent in purely Markovian dynamics. We analyze the
range of validity of this result and identify two relevant time
scales. Finally, we extend the result to more general codes and
argue that the performance of continuous error correction
will exhibit the same qualitative characteristics.
\end{abstract}

\maketitle


\section{Introduction}

Reliable information processing requires the ability to store and
manipulate information with practically negligible loss.
Information carriers, however, constantly interact with
their surroundings, which poses the risk of information being
irreversibly dissipated. This problem is of particular
significance in the case of quantum information, due to the
inherent fragility of quantum superpositions in the presence of
external interactions. Such interactions can quickly lead to
entanglement between the system of interest and its environment,
effectively resulting in the loss of information. This process, known
as decoherence, is a major obstacle in the construction of
large-scale quantum information devices, since as quantum systems
grow in size, they also become increasingly difficult to isolate
from their environment.

Even though decoherence may seem to be a fundamental difficulty,
the development of the theory of quantum fault tolerance
\cite{Sho96, ABO98, Kit97, KLZ98, Got97} has shown that it is
possible in principle to implement reliable quantum information
processing with systems of any size. As long as the error rate per
information unit per time step is kept below a certain threshold,
quantum information can be processed with an arbitrarily small
error. This result is based on the idea of quantum error
correction \cite{Shor95, Steane96, Got97}, where the quantum state
of a single information unit, say a qubit, is encoded in the state
of a larger number of qubits. The encoding is such that if a
single qubit in the code undergoes an error, the original state
can be recovered by applying an appropriate measurement on the
codeword followed by a correcting operation. The success of this
scheme depends on the assumption that individual qubits undergo
independent errors with small probability, and thus that errors on
multiple qubits have probabilities of higher order. This technique
can be extended to multi-qubit errors by constructing more
complicated codes or by concatenation \cite{KL96}.

\subsection{Continuous quantum error correction}

In general, error probabilities increase with time. No matter how
complicated a code or how many levels of concatenation are involved,
the probability of uncorrectable errors is never truly zero, and if the system
is exposed to noise for a sufficiently long time the weight of
uncorrectable errors can accumulate.
To combat this, error correction must be applied
repeatedly and sufficiently often. If one assumes that the time
for an error-correcting operation is small compared to other
relevant time scales of the system, error-correcting operations
can be considered instantaneous. Then the scenario of repeated
error correction leads to a discrete evolution which often may be
difficult to describe.  To study the evolution of a system
in the limit of frequently applied instantaneous error correction, Paz and
Zurek proposed to describe error correction as a continuous
quantum jump process \cite{PZ98}. In this model, the infinitesimal
error-correcting transformation that the density matrix of the
encoded system undergoes during a time step $dt$ is
\begin{equation}\label{basicequation}
\rho\rightarrow (1-\kappa dt)\rho + \kappa dt \Phi(\rho),
\end{equation}
where $\Phi(\rho)$ is the completely positive trace-preserving
(CPTP) map describing a full error-correcting operation, and
$\kappa$ is the error-correction rate. The full error-correcting
operation $\Phi(\rho)$ consists of a syndrome detection, followed
(if necessary) by a
unitary correction operation conditioned on the syndrome.

Consider, for example, the three-qubit bit-flip code whose purpose
is to protect an unknown qubit state from bit-flip (Pauli $X$) errors. The
code space is spanned by $|\overline{0}\rangle = |000\rangle$ and
$|\overline{1}\rangle = |111\rangle$, and the stabilizer
generators are $ZZI$ and $IZZ$.  Here by $X$, $Y$, $Z$ and $I$ we denote
the usual Pauli operators and the identity, respectively,
and a string of three operators represents
the tensor product of operators on each of the three qubits.
The standard error-correction procedure
involves a measurement of the stabilizer generators, which
projects the state onto one of the subspaces spanned by
$|000\rangle$ and $|111\rangle$, $|100\rangle$ and $|011\rangle$,
$|010\rangle$ and $|101\rangle$, or $|001\rangle$ and
$|110\rangle$; the outcome of these measurements is the
error syndrome. Assuming that the probability for two- or
three-qubit errors is negligible, then with high probability the
result of this measurement is either the original state with
no errors, or with a single $X$ error on the first, the second, or the
third qubit. Depending on the outcome, one then applies an $X$
gate to the erroneous qubit and transforms the state back to the
original one. The CPTP map $\Phi(\rho)$ for this code can be
written explicitly as
\begin{equation}
\begin{split}
\Phi(\rho) = \left(|000\rangle \langle000| + |111\rangle \langle111| \right) \rho
\left(|000\rangle \langle000| + |111\rangle \langle111| \right) \\
+ \left(|000\rangle \langle100| + |111\rangle \langle011| \right) \rho
\left(|100\rangle \langle000| + |011\rangle \langle111| \right) \\
+ \left(|000\rangle \langle010| + |111\rangle \langle101| \right)\rho
\left(|010\rangle \langle000| + |101\rangle \langle111| \right) \\
+ \left(|000\rangle\langle001| + |111\rangle \langle110| \right)\rho
\left(|001\rangle \langle000| + |110\rangle \langle111| \right)
\label{strongmap}
\end{split}
\end{equation}

The quantum-jump process \eqref{basicequation} can be viewed as a
smoothed version of the discrete scenario of repeated error
correction, in which instantaneous full error-correcting
operations are applied at random times with rate $\kappa$. It can
also be looked upon as arising from a continuous sequence of
infinitesimal CPTP maps of the type \eqref{basicequation}. In
practice, such a weak map is never truly infinitesimal, but rather
has the form
\begin{equation} \rho
\rightarrow (1-\varepsilon)\rho + \varepsilon
\Phi(\rho),\label{wm}
\end{equation}
where $\varepsilon \ll 1$ is a small but finite parameter, and the weak operation takes a
small but nonzero time $\tau_c$. For
times $t$ much greater than $\tau_c$ ($\tau_c\ll t$), the
weak error-correcting map (\ref{wm}) is well approximated by the infinitesimal
form \eqref{basicequation}, where the rate of error correction is
\begin{equation}
\kappa = \varepsilon /\tau_c. \label{tauc}
\end{equation}
A weak map of the form \eqref{wm} could be implemented, for
example, by a weak coupling between the system and an ancilla via
an appropriate Hamiltonian, followed by discarding the ancilla. A
closely related scenario, where the ancilla is continuously cooled
in order to reset it to its initial state, was studied in
\cite{SarMil05}.

Another way of implementing the weak map is
via weak measurements followed by weak unitaries dependent on the
outcome. The corresponding weak measurements, however, are not
weak versions of the strong measurements for syndrome detection;
they are in a different basis \cite{OBinprep}. They can be
regarded as weak versions of a {\it different} set of strong
measurements which, when followed by an appropriate unitary, yield
the same map $\Phi(\rho)$ on average. Thus, the workings of continuous
error correction, when it is driven by weak measurements, does not translate
directly into the error syndrome detection and correction of the
standard paradigm. In this sense, the continuous approach can be
regarded as a different paradigm for error correction---one based
on weak measurements and weak unitary operations. The idea of
using continuous weak measurements and unitary operations for
error correction has been explored in the context of different
heuristic schemes \cite{ADL02, SarMil05g}, some of which are based
on a direct ``continuization'' of the syndrome measurements. In this
paper we consider continuous error correction of the type
given by Eq.~\eqref{basicequation}.

\subsection{Markovian decoherence}

So far, continuous quantum error correction has been studied only
for Markovian error models. The Markovian approximation describes
situations where the bath-correlation times are much shorter than
any characteristic time scale of the system \cite{BrePet02}. In
this limit, the dynamics can be described by a semi-group master
equation in the Lindblad form \cite{Lin76}:
\begin{equation}
\frac{d\rho}{dt}=L(\rho)\equiv-i[H,\rho]+\frac{1}{2}\underset{j}{\sum}\lambda_j(2L_j\rho
L_j^{\dagger}-L_j^{\dagger}L_j\rho-\rho L_j^{\dagger}L_j).
\end{equation}
Here $H$ is the system Hamiltonian and the $\{L_j\}$ are suitably
normalized Lindblad operators describing different error channels
with decoherence rates $\lambda_j$. For example, the Liouvillian
\begin{equation}
L(\rho)= \underset{j}{\sum}\lambda_j(X_j\rho X_j - \rho),
\label{Lbitflip}
\end{equation}
where $X_j$ denotes a local bit-flip operator acting on the $j$-th
qubit, describes independent Markovian bit-flip errors.

For a system undergoing Markovian decoherence and error correction
of the type \eqref{basicequation}, the evolution is given by the
equation
\begin{equation}
\frac{d\rho}{dt}=L(\rho)+\kappa\Gamma(\rho),\label{errorcorrectionequation}
\end{equation}
where $\Gamma(\rho)=\Phi(\rho)-\rho$. In \cite{PZ98}, Paz and
Zurek showed that if the set of errors $\{L_j\}$ are correctable by
the code, in the limit of infinite error-correction rate (strong
error-correcting operations applied continuously often) the state
of the system freezes and is protected from errors at all times.
The effect of freezing can be understood by noticing that the
transformation arising from decoherence during a short time step
$\Delta t$, is
\begin{equation}
\rho\rightarrow \rho + L(\rho)\Delta t +\textit{O}(\Delta t^2),
\end{equation}
i.e., the weight of correctable errors emerging during this time
interval is proportional to $\Delta t$, whereas uncorrectable
errors (e.g. multi-qubit bit flips in the case of the three-qubit
bit-flip code) are of order $\textit{O}(\Delta t^2)$. Thus, if
errors are constantly corrected, in the limit $\Delta t
\rightarrow 0$ uncorrectable errors cannot accumulate, and the
evolution stops.

\subsection{The Zeno effect. Error correction versus error prevention}

The effect of ``freezing'' in continuous error correction strongly
resembles the quantum Zeno effect \cite{MisSud77}, in which
frequent measurements slow down the evolution of a system,
freezing the state in the limit where they are applied
continuously. The Zeno effect arises when the system and its
environment are initially decoupled and they undergo a
Hamiltonian-driven evolution, which leads to a quadratic change
with time of the state during the initial moments \cite{NNP96}
(the so called Zeno regime). Let the initial state of the system
plus the bath be $\rho_{SB}(0)=|0\rangle \langle
0|_S\otimes\rho_B(0)$. For small times, the fidelity of the
system's density matrix with the initial state
$\alpha(t)=\textrm{Tr}\left\{\left(|0\rangle\langle0|_S\otimes
I_B\right)\rho_{SB}(t)\right\}$ can be approximated as
\begin{equation}
\alpha(t)= 1-C t^2+\textit{O}(t^3).\label{Zeno}
\end{equation}
In terms of the Hamiltonian $H_{SB}$ acting on the entire system,
the coefficient $C$ is
\begin{equation}
C = \textrm{Tr}\left\{H_{SB}^2\left(|0\rangle\langle0|_S\otimes\rho_B(0)\right)\right\}
- \textrm{Tr}\left\{H_{SB}\left(|0\rangle\langle0|_S\otimes I_B\right)
H_{SB} \left(|0\rangle\langle0|_S\otimes \rho_B(0)\right)\right\}.
\label{C}
\end{equation}
According to Eq. \eqref{Zeno}, if after a short time step $\Delta
t$ the system is measured in an orthogonal basis which includes
the initial state $|0\rangle$, the probability to find the system
in a state other than the initial state is of order
$\textit{O}(\Delta t^2)$. Thus if the state is continuously
measured ($\Delta t \rightarrow 0$), this prevents the system from
evolving.

It has been proposed to utilize the quantum Zeno effect in schemes
for error prevention \cite{Zur84, BBDEJM97, VGW96}, in which an
unknown encoded state is prevented from errors simply by frequent
measurements which keep it inside the code space. The approach is
similar to error correction in that the errors for which the code
is designed send a codeword to a space orthogonal to the code
space. The difference is that different errors need not be distinguishable,
since the procedure does not involve {\it correction} of errors, but their prevention.
In \cite{VGW96} it was shown that with this approach it is possible
to use codes of smaller redundancy than those needed for error
correction and a four-qubit encoding of a qubit was proposed,
which is capable of preventing arbitrary independent errors
arising from Hamiltonian interactions. The possibility of this
approach implicitly assumes the existence of a Zeno regime, and fails if
we assume Markovian decoherence for all times. This is because the
probability of errors emerging during a time step $dt$ in a Markovian model is
proportional to $dt$  (rather than $dt^2$), and hence errors will
accumulate with time if not corrected.

From the above observations we see that error {\it correction} is capable
of achieving results in noise regimes where error {\it prevention}
fails. Of course, this advantage is at the expense of a more
complicated procedure---in addition to the measurements used in
error prevention, error correction involves unitary correction operations,
and in general requires codes with higher redundancy.
At the same time, we see that in the Zeno regime it is possible to
reduce decoherence using weaker resources than those needed in the
case of Markovian noise. This suggests that in this regime error
correction may exhibit higher performance than it does
for Markovian decoherence.

\subsection{Non-Markovian decoherence}

Markovian decoherence is an approximation valid for times much
larger than the memory of the environment. In many situations of
practical significance, however, the memory of the environment
cannot be neglected and the evolution is highly non-Markovian
\cite{BrePet02,QWJ97, BBP04, KORL07}. Furthermore, no evolution is
strictly Markovian, and for a system initially decoupled from its
environment a Zeno regime is always present, short though it may be
\cite{NNP96}. If the time resolution of
error-correcting operations is high enough so that they ``see'' the
Zeno regime, this could give rise to different behavior.

The existence of a Zeno regime is not the only interesting feature
of non-Markovian decoherence. The mechanism by which errors
accumulate in a general Hamiltonian interaction with the
environment may differ significantly from the Markovian case,
since the system may develop nontrivial correlations with the
environment. For example, imagine that some time after the initial
encoding of a system, a strong error-correcting operation is
applied. This brings the state inside the code space, but the
state contains a nonzero portion of errors non-distinguishable by
the code. Thus the new state is mixed and is generally correlated
with the environment. A subsequent error-correcting operation can
only aim at correcting errors arising after this point, since the
errors already present inside the code space are in principle
uncorrectable. Subsequent errors on the density matrix, however,
may not be completely positive due to the correlations with the
environment.

Nevertheless, it follows from a result in
\cite{ShaLid06} that an error-correction procedure which is
capable of correcting a certain class of completely positive (CP)
maps, can also correct any linear noise map whose operator
elements can be expressed as linear combinations of the operator
elements in a correctable CP map. This implies, in particular,
that an error-correction procedure that can correct arbitrary
single-qubit CP maps can correct arbitrary
single-qubit linear maps. The effects of system-environment
correlations in non-Markovian error models have also been studied
from the perspective of fault tolerance, and it
has been shown that the threshold theorem can be extended to
various types of non-Markovian noise \cite{TB05, AGP06, AKP06}.

Another important difference from the Markovian case is that error
correction and the effective noise on the reduced density matrix
of the system cannot be treated as independent processes. One
could derive an equation for the effective evolution of the system
alone subject to interaction with the environment, like the
Nakajima-Zwanzig \cite{Nak58, Zwa60} or the time-convolutionless
(TCL) \cite{Shibata77, ShiAri80} master equations, but the
generator of transformations at a given moment in general will
depend (implicitly or explicitly) on the entire history up to this
moment. Therefore, adding error correction can nontrivially affect
the effective error model. This means that in studying the
performance of continuous error correction one either has to
derive an equation for the effective evolution of the encoded
system, taking into account error correction from the very
beginning, or one has to look at the evolution of the entire
system---including the bath---where the error generator and the
generator of error correction can be considered independent. In
the latter case, for sufficiently small $\tau_c$, the evolution of
the entire system including the bath can be described by
\begin{equation}
\frac{d \rho}{dt}=-i[H, \rho] + \kappa \Gamma(\rho),
\label{NMerrorcorrectionequation}
\end{equation}
where $\rho$ is the density matrix of the system plus bath,
$H$ is the total Hamiltonian, and the error-correction generator
$\Gamma$ acts locally on the encoded system. In this paper, we
take this approach for a sufficiently simple bath model which
allows us to find a solution for the evolution of the entire
system.

\subsection{Plan of this paper}

The rest of the paper is organized as follows. To develop
understanding of the workings of continuous error correction, in
Sec. II we look at a simple example:  an error-correction code
consisting of only one qubit which aims at protecting a known
state. We discuss the difference in performance for Markovian and
non-Markovian decoherence, and argue the implications it has for
the case of multi-qubit codes. In Sec. III, we study the
three-qubit bit-flip code. We first review the performance of
continuous error correction in the case of Markovian bit-flip
decoherence, which was first studied in \cite{PZ98}. We then
consider a non-Markovian model, where each qubit in the code is
coupled to an independent bath qubit. This model is sufficiently
simple so that we can solve for its evolution analytically.  In
the limit of large error-correction rates, the effective evolution
approaches the evolution of a single qubit without error
correction, but the coupling strength is now decreased by a factor
which scales quadratically with the error-correction rate. This is
opposed to the case of Markovian decoherence, where the same
factor scales linearly with the rate of error-correction. In Sec.
IV, we show that the quadratic enhancement in the performance over
the case of Markovian noise can be attributed to the presence of a
Zeno regime and argue that for general stabilizer codes and
independent errors, the performance of continuous error correction
would exhibit the same qualitative characteristics. In Sec. V, we
conclude.

\section{The single-qubit code}

Consider the problem of protecting a qubit in state $|0\rangle$
from bit-flip errors. This problem can be regarded as a trivial
example of a stabilizer code, where the code space is spanned by
$|0\rangle$ and its stabilizer is $Z$. Let us consider the
Markovian bit-flip model first. The evolution of the state subject
to bit-flip errors and error correction is described by Eq.
\eqref{errorcorrectionequation} with
\begin{equation}
L(\rho)=\lambda( X \rho X - \rho),
\label{bitflipgen}
\end{equation}
and
\begin{equation}
\Gamma(\rho)=|0\rangle \langle 0| \rho |0\rangle \langle0|
+ |0\rangle \langle 1|\rho |1\rangle \langle 0| - \rho.
\label{ECgen}
\end{equation}
If the state lies on the z-axis of the Bloch sphere, it will never
leave it, since both the noise generator \eqref{bitflipgen} and
the error-correction generator \eqref{ECgen} keep it on the axis.
We will take the qubit to be initially in the desired state
$|0\rangle$, and therefore at any later moment it will have the form
$\rho (t) = \alpha(t) |0\rangle\langle 0|+(1-\alpha(t))|1\rangle\langle 1|$,
$\alpha(t) \in [0,1]$. The coefficient $\alpha(t)$ has the interpretation
of a fidelity with the trivial code
space spanned by $|0\rangle$. For an infinitesimal time step $dt$,
the effect of the noise is to decrease $\alpha(t)$ by the amount
$\lambda (2\alpha(t)-1) dt$ and that of the correcting operation
is to increase it by $\kappa (1-\alpha(t)) dt$. The net evolution is
then described by
\begin{equation}
\label{equation1}
\frac{d\alpha(t)}{dt}=-(\kappa+2\lambda)\alpha(t)+(\kappa + \lambda).
\end{equation}
The solution is
\begin{equation}
\alpha(t)=(1-\alpha_*^{\rm M})e^{-(\kappa+2\lambda)t}+\alpha_*^{\rm M},
\label{MSQS}
\end{equation}
where
\begin{equation}
\alpha_*^{\rm M}=1-\frac{1}{2+r},
\label{attractor}
\end{equation}
and $r=\kappa/\lambda$ is the ratio between the rate of error
correction and the rate of decoherence. We see that the fidelity
decays, but it is confined above its asymptotic value
$\alpha_*^{\rm M}$, which can be made arbitrarily close to 1 for a
sufficiently large $r$.

Now let us consider a non-Markovian error model. We choose the
simple scenario where the system is coupled to a single bath qubit
via the Hamiltonian
\begin{equation}
H=\gamma X\otimes X,
\end{equation}
where $\gamma$ is the coupling strength. This can be a good
approximation for situations in which the coupling to a single
spin from the bath dominates over other interactions
\cite{KORL07}.

We will assume that the bath qubit is initially in the maximally
mixed state, which can be thought of as an equilibrium state at
high temperature. From Eq. \eqref{NMerrorcorrectionequation} one
can verify that if the system is initially in the state
$|0\rangle$, the state of the system plus the bath at any moment
will have the form
\begin{eqnarray}
\rho(t) = \left(\alpha (t) |0\rangle \langle 0| + (1-\alpha(t))|1\rangle \langle 1|\right)\otimes \frac{I}{2}
- \beta(t)Y \otimes \frac{X}{2}.
\end{eqnarray}
In the tensor product, the first operator belongs to the Hilbert
space of the system and the second to the Hilbert space of the
bath. We have $\alpha(t) \in [0,1]$, and
$|\beta(t)|\le\sqrt{\alpha(t)(1-\alpha(t))}, \beta(t)\in R$. The
reduced density matrix of the system has the same form as the one
for the Markovian case. The traceless term proportional to $\beta(t)$ can be
thought of as a ``hidden'' part, which nevertheless plays an
important role in the error-creation process, since errors can be
thought of as being transferred to the ``visible'' part from the
``hidden'' part (and vice versa). This can be seen from the fact
that during an infinitesimal time step $dt$, the Hamiltonian
changes the parameters $\alpha$ and $\beta$ as follows:
\begin{gather}
\alpha\rightarrow \alpha-2\beta \gamma dt ,\notag\\
\beta \rightarrow \beta +(2\alpha-1)\gamma dt .\label{sqe}
\end{gather}
The effect of an infinitesimal error-correcting operation is
\begin{gather}
\alpha \rightarrow \alpha + (1-\alpha)\kappa dt,\notag\\
\beta\rightarrow \beta-\beta\kappa dt.
\end{gather}
Note that the hidden part is also being acted upon. Putting it all
together, we get the system of equations
\begin{gather}
\frac{d \alpha(t)}{dt}=\kappa(1-\alpha(t))-2\gamma \beta(t),\notag\\
\frac{d
\beta(t)}{dt}=\gamma(2\alpha-1)-\kappa\beta(t)\label{equation2}.
\end{gather}
The solution for the fidelity $\alpha(t)$ is
\begin{gather}
\alpha(t) = \frac{2\gamma^2 + \kappa^2}{4\gamma^2+\kappa^2}
+ e^{-\kappa t}\left(\frac{\kappa\gamma}{4\gamma^2+\kappa^2} \sin{2\gamma t}
+ \frac{2\gamma^2}{4\gamma^2+\kappa^2}\cos{2\gamma t}\right).
\label{singlequbitsolution}
\end{gather}
We see that as time increases, the fidelity stabilizes at the
value
\begin{equation}
\alpha_*^{\rm NM}= \frac{2+R^2}{4+R^2}=1-\frac{2}{4+R^2},
\end{equation}
where $R=\kappa/\gamma$ is the ratio between the error-correction
rate and the coupling strength. In Fig. 1 we have plotted the
fidelity as a function of the dimensionless parameter $\gamma t$
for three different values of $R$. For error-correction rates
comparable to the coupling strength ($R=1$), the fidelity
undergoes a few partial recurrences before it stabilizes close to
$\alpha_*^{\rm NM}$. For larger $R=2$, however, the oscillations are
already heavily damped and for $R=5$ the fidelity seems confined
above $\alpha_*^{\rm NM}$. As $R$ increases, the evolution becomes
closer to a decay like the one in the Markovian case.

\begin{figure}[h]
\begin{center}
\includegraphics[width=4in]{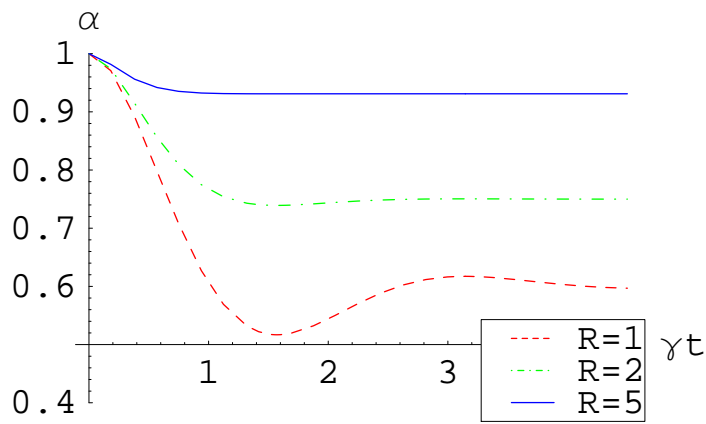}
\caption{(Color online)  Fidelity of the single-qubit code with
continuous bit-flip errors and correction, as a function of
dimensionless time $\gamma t$, for three different values of the
ratio $R=\kappa/\gamma$.} \label{fig1}
\end{center}
\end{figure}

A remarkable difference, however, is that the asymptotic weight
outside the code space ($1-\alpha_*^{\rm NM}$) decreases with
$\kappa$ as $1/\kappa^2$, whereas in the Markovian case the same
quantity decreases as $1/\kappa$. The asymptotic value can be
obtained as an equilibrium point at which the infinitesimal weight
flowing out of the code space during a time step $dt$ is equal to
the weight flowing into it. The latter corresponds to vanishing
right-hand sides in Eqs. \eqref{equation1} and \eqref{equation2}.
In Sec. IV, we will show that the difference in the equilibrium
code-space fidelity for the two different types of decoherence
arises from the difference in the corresponding evolutions during
initial times.

For multi-qubit codes, error correction cannot preserve a high
fidelity with the initial codeword for all times, because there
will be multi-qubit errors that can lead to errors within the code
space itself. But it is natural to expect that the code-space
fidelity can be kept above a certain value, since the effect of
the error-correcting map \eqref{basicequation} is to oppose its
decrease. If similarly to the single-qubit code there is a
quadratic difference in the code-space fidelity for the cases of
Markovian and non-Markovian decoherence, this could lead to a
different performance of the error-correction scheme with respect
to the rate of accumulation of uncorrectable errors inside the
code space. This is because multi-qubit errors that can lead to
transformations entirely within the code space during a time step
$dt$ are of order $\textit{O}(dt^2)$. This means that if the state
is kept constantly inside the code space (as in the limit of an
infinite error-correction rate), uncorrectable errors will never
develop. But if there is a finite nonzero portion of correctable
errors, by the error mechanism it will give rise to errors not
distinguishable or misinterpreted by the code. Therefore, the
weight outside the code space can be thought of as responsible for
the accumulation of uncorrectable errors, and consequently a
difference in its magnitude may lead to a difference in the
overall performance. In the following sections we will see that
this is indeed the case.

\section{The three-qubit bit-flip code}

\subsection{A Markovian error model}

Even though the three-qubit bit-flip code can correct only
bit-flip errors, it captures most of the important characteristics
of nontrivial stabilizer codes. Before we look at a non-Markovian
model, we will review the Markovian case which was studied in
\cite{PZ98}. Let the system decohere through identical independent
bit-flip channels, i.e., $L(\rho)$ is of the form \eqref{Lbitflip}
with $\lambda_1=\lambda_2=\lambda_3=\lambda$. Then one can verify
that the density matrix at any moment can be written as
\begin{equation}
\rho(t) = a(t)\rho(0)+b(t)\rho_{1}+c(t)\rho_{2}+d(t)\rho_{3},
\label{rhooft}
\end{equation}
where
\begin{gather}
\rho_{1}=\frac{1}{3}(X_1\rho(0)X_1+X_2\rho(0)X_2 +
X_3\rho(0)X_3),\notag\\
\rho_{2}= \frac{1}{3}(X_1X_2\rho(0)X_1X_2+
X_2X_3\rho(0)X_2X_3+X_1X_3\rho(0)X_1X_3),\\
\rho_{3} = X_1X_2X_3\rho(0) X_1X_2X_3,\notag
\end{gather}
are equally-weighted mixtures of single-qubit, two-qubit and
three-qubit errors on the original state.

The effect of decoherence for a single time step $dt$ is
equivalent to the following transformation of the coefficients in
Eq. \eqref{rhooft}:
\begin{equation}
\begin{split}
a\rightarrow a - 3a \lambda dt + b \lambda dt,\\
b\rightarrow b + 3a \lambda dt - 3 b \lambda dt + 2 c \lambda dt,\\
c\rightarrow c +2b \lambda dt - 3c\lambda dt+3d\lambda dt,\\
d\rightarrow d +c \lambda dt -3d \lambda dt.
\label{decohtransform}
\end{split}
\end{equation}
If the system is initially inside the code space, combining Eq.
\eqref{decohtransform} with the effect of the weak
error-correcting map $\rho\rightarrow (1-\kappa dt)\rho + \kappa
dt \Phi(\rho)$, where $\Phi(\rho)$ is given in Eq.
\eqref{strongmap}, yields the following system of first-order
linear differential equations for the evolution of the system
subject to decoherence plus error correction:
\begin{equation}
\begin{split}
\frac{da(t)}{dt} = -3\lambda a(t) + (\lambda+\kappa)b(t),\\
\frac{db(t)}{dt} = 3\lambda a(t) - (3\lambda+\kappa)b(t) + 2 \lambda c(t),\\
\frac{dc(t)}{dt} = 2\lambda b(t) - (3\lambda+\kappa)c(t) + 3 \lambda d(t),\\
\frac{dd(t)}{dt} = (\lambda+\kappa)c(t)-3\lambda d(t).
\label{equations}
\end{split}
\end{equation}
The exact solution has been found in \cite{PZ98}. Here we just
note that for the initial conditions $a(0)=1, b(0)=c(0)=d(0)=0$,
the exact solution for the weight outside the code space is
\begin{equation}
b(t)+c(t)=\frac{3}{4+r}(1-e^{-(4+r)\lambda t}),
\end{equation}
where $r=\kappa/\lambda$. We see that similarly to what we
obtained for the trivial code in the previous section, the weight
outside the code space quickly decays to its asymptotic value
$\frac{3}{4+r}$ which scales as $1/r$. But note that here the
asymptotic value is roughly three times greater than that for the
single-qubit model. This corresponds to the fact that there are
three single-qubit channels. More precisely, it can be verified
that if for a given $\kappa$ the uncorrected weight by the
single-qubit scheme is small, then the uncorrected weight by a
multi-qubit code using the same $\kappa$ and the same kind of
decoherence for each qubit scales approximately linearly with the
number of qubits \cite{OBinprep}. Similarly, the ratio $r$
required to preserve a given overlap with the code space scales
linearly with the number of qubits in the code.

The most important difference from the single-qubit model is that
in this model there are uncorrectable errors that cause a decay
of the state's fidelity {\it inside} the code space. Due to the finiteness of the
resources employed by our scheme, there always remains a nonzero
portion of the state outside the code space, which gives rise to
uncorrectable three-qubit errors. To understand how the state
decays inside the code space, we ignore the terms of the order of
the weight outside the code space in the exact solution. We
obtain:
\begin{equation}
a(t)\approx \frac{1+e^{-\frac{6}{r}2\lambda t}}{2} \approx 1 -
d(t),
\end{equation}
\begin{equation}
b(t) \approx c(t) \approx 0.
\end{equation}
Comparing this solution to the expression for the fidelity of a
single decaying qubit without error correction---which can be seen
from Eq. \eqref{MSQS} for $\kappa=0$---we see that the encoded
qubit decays roughly as if subject to bit-flip decoherence with
rate $6\lambda/r$. Therefore, for large $r$ this error-correction
scheme can reduce the rate of decoherence approximately $r/6$
times. In the limit $r \rightarrow \infty$, it leads to perfect
protection of the state for all times.

\subsection{A non-Markovian error model}

We consider a model where each qubit independently undergoes the
same kind of non-Markovian decoherence as the one we studied for
the single-qubit code. Here the system we look at consists of six
qubits - three for the codeword and three for the environment. We
assume that all system qubits are coupled to their corresponding
environment qubits with the same coupling strength, i.e., the
Hamiltonian is
\begin{equation}
H=\gamma\overset{3}{\underset{i=1}{\sum}}X^S_i\otimes
X^B_i,\label{Hamiltonian}
\end{equation}
where the operators $X^S$ act on the system qubits and $X^B$ act
on the corresponding bath qubits. The subscripts label the
particular qubit on which they act. Obviously, the types of effective
single-qubit errors on the density matrix of the system that can
result from this Hamiltonian at any time, whether they are CP or not,
will have operator elements which are linear combinations of $I$
and $X^S$, i.e., they are correctable by the procedure according
to \cite{ShaLid06}. Considering the forms of the Hamiltonian
\eqref{Hamiltonian} and the error-correcting map
\eqref{strongmap}, one can see that the density matrix of the
entire system at any moment is a linear combination of terms of
the following type:
\begin{equation}
\varrho_{lmn,pqr}\equiv X_1^lX_2^mX_3^n\rho(0)
X_1^pX_2^qX_3^r\otimes \frac{X_1^{l+p}}{2}\otimes
\frac{X_2^{m+q}}{2} \otimes \frac{X_3^{n+r}}{2}.
\end{equation}
Here the first term in the tensor product refers to the Hilbert
space of the system, and the following three refer to the Hilbert
spaces of the bath qubits that couple to the first, second and
third qubits from the code, respectively. The powers
$l,m,n,p,q,r$ take values $0$ and $1$ in all possible
combinations, and $X^1=X$, $X^0=X^2=I$.  Note that
$\varrho_{lmn,pqr}$ should not be mistaken for the components of
the density matrix in the computational basis. Collecting these together, we
can write the density matrix in the form
\begin{eqnarray}
\rho(t)&=&\underset{l,m,n,p,q,r}{\sum}(-i)^{l+m+n}(i)^{p+q+r}C_{lmn,pqr}(t)\times
\varrho_{lmn,pqr},\label{fullDM}
\end{eqnarray}
where the coefficients $C_{lmn,pqr}(t)$ are real. The coefficient
$C_{000,000}$ is less than or equal to the codeword fidelity
(with equality when $\rho(0)=|\bar{0}\rangle\langle \bar{0}|$ or
$\rho(0)=|\bar{1}\rangle\langle \bar{1}|$). Since the scheme is intended
to protect an unknown codeword, we are interested in its worst-case
performance; we will therefore use $C_{000,000}$ as a lower bound
on the codeword fidelity.

Using the symmetry with respect to permutations of the different
system-bath pairs of qubits and the Hermiticity of the density
matrix, we can reduce the description of the evolution to a system of
equations for only $13$ of the $64$ coefficients.  (In fact, $12$ coefficients
are sufficient if we invoke the normalization condition $\textrm{Tr}\rho=1$, but we
have found it more convenient to work with $13$.) The equations are linear, and we
write them as a single 13-dimensional vector equation:
\begin{equation}
\setcounter{MaxMatrixCols}{13} \frac{d}{dt}\begin{bmatrix}
C_{000,000}\\
C_{100,000}\\
C_{110,000}\\
C_{100,010}\\
C_{100,100}\\
C_{110,001}\\
C_{111,000}\\
C_{110,100}\\
C_{110,110}\\
C_{110,011}\\
C_{111,100}\\
C_{111,110}\\
C_{111,111}
\end{bmatrix}=\gamma
\begin{bmatrix}
0&-6&0&0&3R&0&0&0&0&0&0&0&0\\
1&-R&-2&-2&-1&0&0&0&0&0&0&0&0\\
0&2&-R&0&0&-1&-1&-2&0&0&0&0&0\\
0&2&0&-R&0&-2&0&-2&0&0&0&0&0\\
0&2&0&0&-R&0&0&-4&0&0&0&0&0\\
0&0&1&2&0&-R&0&0&0&-2&-1&0&0\\
0&0&3&0&0&-3R&0&0&0&0&-3&0&0\\
0&0&1&1&1&0&0&-R&-1&-1&-1&0&0\\
0&0&0&0&0&0&0&4&-R&0&0&-2&0\\
0&0&0&0&0&2&0&2&0&-R&0&-2&0\\
0&0&0&0&0&1&1&2&0&0&-R&-2&0\\
0&0&0&0&0&0&0&0&1&2&2&-R&-1\\
0&0&0&0&0&0&0&0&3R&0&0&6&0
\end{bmatrix}\cdot
\begin{bmatrix}
C_{000,000}\\
C_{100,000}\\
C_{110,000}\\
C_{100,010}\\
C_{100,100}\\
C_{110,001}\\
C_{111,000}\\
C_{110,100}\\
C_{110,110}\\
C_{110,011}\\
C_{111,100}\\
C_{111,110}\\
C_{111,111}
\end{bmatrix}
\label{NMsystem1}
\end{equation}
where $R=\kappa/\gamma$.  Each nonzero component in this matrix
represents an allowed transition process for the quantum states; these
transitions can be driven either by the decoherence process or the
continuous error-correction process.  We plot these allowed
transitions in Fig.~2.

\begin{figure}[htbp]
\includegraphics[width=6in]{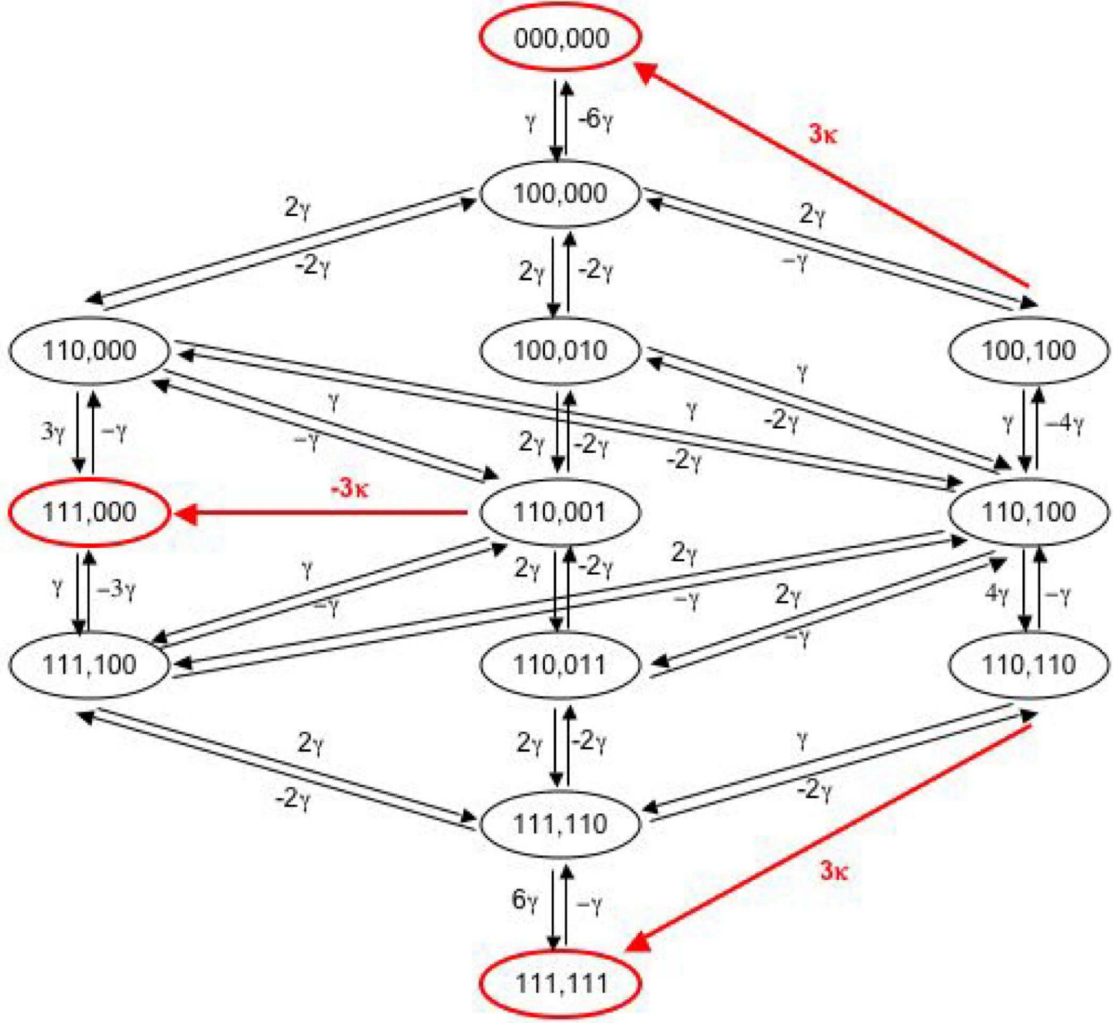}
\caption{(Color online)  These are the allowed transitions between
the different components of the system (\ref{NMsystem1}) and their
rates, arising from both the decoherence (bit-flip) process (with
rate $\gamma$ and the continuous error-correction process (with
rate $\kappa$).  Online, the transitions due to decoherence are
black, and the transitions due to error correction are red.}
\label{fig2}
\end{figure}

We can use the symmetries of the process to recover the 64
coefficients of the full state.  Each of
the 13 coefficients represents a set of coefficients having the
same number of $1$s on the left and the same number of $1$s on the
right, as well as the same number of places which have $1$ on both
sides.  All such coefficients are equal at all times.
For example, the coefficient $C_{110,011}$ is equal to all
coefficients with two $1$s on the left, two $1$s on the right and
exactly one place with $1$ on both sides; there are exactly six such
coefficients:
\[
C_{110,011} = C_{110,101} = C_{101,011} = C_{101,110} = C_{011,110} = C_{011,101} .
\]
In determining the transfer rate from one coefficient to another
in Fig.~2, one has to take into account the number of different
coefficients of the first type which can make a transition to a
coefficient of the second type of order $dt$ according to Eq.
\eqref{NMerrorcorrectionequation}. The sign of the flow is
determined from the phases in front of the coefficients in Eq.
\eqref{fullDM}.

The eigenvalues of the matrix in Eq. \eqref{NMsystem1} up to the
first two lowest orders in $1/\kappa$ are presented in Table I.

\begin{table}[htdp]
\caption{Eigenvalues of the matrix}
\begin{center}
\begin{tabular}{|c|c|}
\hline Eigenvalues \\ \hline $\lambda_0 = 0$   \\ \hline
$\lambda_{1,2} = -\kappa$ \\ \hline $\lambda_{3,4} =  - \kappa\pm
i 2\gamma$ \\ \hline $\lambda_{5,6} =  - \kappa \pm i 4\gamma$ \\
\hline $\lambda_{7,8} = -\kappa\pm
i(\sqrt{13}+3)\gamma+\textit{O}(1/\kappa)$ \\ \hline
$\lambda_{9,10} = -\kappa\pm
i(\sqrt{13}-3)\gamma+\textit{O}(1/\kappa)$
\\ \hline $\lambda_{11,12} = \pm i(24/R^2)\gamma  - (144/R^3)
\gamma + \textit{O}(1/\kappa^4)$ \\ \hline
\end{tabular}
\end{center}
\label{eigenvalue_table}
\end{table}%
Obviously all eigenvalues except the first one and the last two
describe fast decays with rates $\sim \kappa$. They correspond to terms
in the solution which will vanish quickly after the beginning of
the evolution. The eigenvalue $0$ corresponds to the asymptotic
($t\rightarrow \infty$) solution, since all other terms will
eventually decay. The last two eigenvalues are those that play the
main role in the evolution on a time scale $t\gg\frac{1}{\kappa}$.
We see that on such a time scale, the solution will contain an
oscillation with an angular frequency approximately equal to
$(24/R^2)\gamma$ which is damped by a decay factor with a
rate of approximately $(144/R^3)\gamma$.  In Fig.~3 we have
plotted the codeword fidelity $C_{000,000}(t)$ as a function of
the dimensionless parameter $\gamma t$ for $R=100$. The graph
indeed represents this type of behavior, except for very short
times after the beginning ($\gamma t \sim 0.1$), where one can see
a fast but small in magnitude decay (Fig. 4). The maximum
magnitude of this quickly decaying term obviously decreases with
$R$, since in the limit of $R\rightarrow \infty$ the fidelity
should remain constantly equal to $1$.

\begin{figure}[h]
\begin{center}
\includegraphics[width=4in]{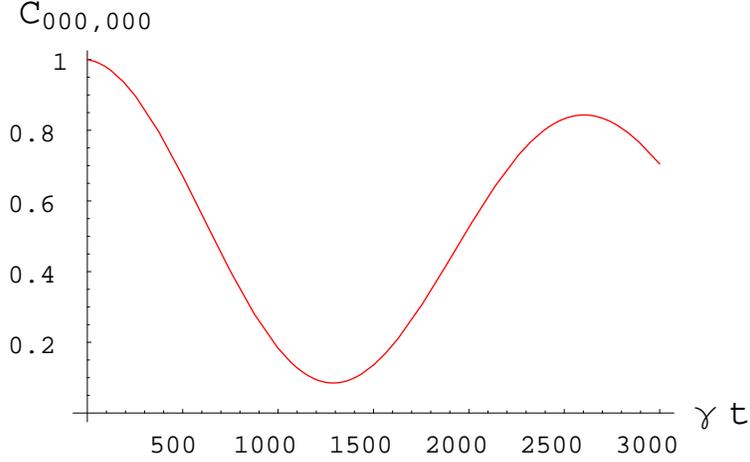}
\caption{(Color online)  Long-time behavior of three-qubit system
with bit-flip noise and continuous error correction.  The ratio of
correction rate to decoherence rate is $R=\kappa/\gamma=100$.}
\label{fig3}
\end{center}
\end{figure}

\begin{figure}[h]
\begin{center}
\includegraphics[width=4in]{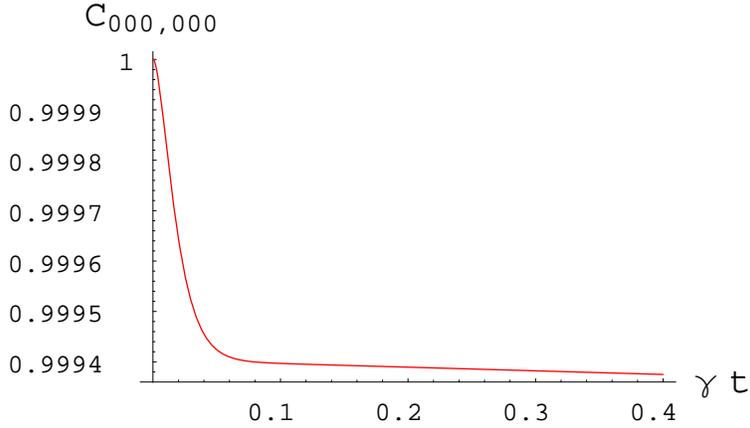}
\caption{(Color online)  Short-time behavior of three-qubit system
with bit-flip noise and continuous error correction.  The ratio of
correction rate to decoherence rate is $R=\kappa/\gamma=100$.}
\label{fig4}
\end{center}
\end{figure}

From the form of the eigenvalues one can see that as $R$
increases, the frequency of the main oscillation decreases as
$1/R^2$ while the rate of decay decreases faster, as $1/R^3$. Thus
in the limit $R\rightarrow \infty$, the evolution approaches an
oscillation with an angular frequency $(24/R^2)\gamma$. (We
formulate this statement more rigorously below.) This is the same
type of evolution as that of a single qubit interacting with its
environment, but the coupling constant is effectively reduced by a
factor of $R^2/12$.

While the coupling constant serves to
characterize the decoherence process in this particular case, this
is not valid in general. To handle the more general situation,
we propose to use the
instantaneous rate of decrease of the codeword fidelity $F_{cw}$
as a measure of the effect of decoherence:
\begin{equation}
\Lambda(F_{cw}(t)) = -\frac{dF_{cw}(t)}{dt}.
\label{errorrate}
\end{equation}
(In the present case, $F_{cw}=C_{000,000}$.)
This quantity does not coincide with the decoherence rate in the
Markovian case (which can be defined naturally from the Lindblad
equation), but it is a good estimate of the rate of loss of
fidelity and can be used for any decoherence model. From now on we
will refer to it simply as an error rate, but we note that there
are other possible definitions of instantaneous error rate
suitable for non-Markovian decoherence, which in general may
depend on the kind of errors they describe. Since the goal of
error correction is to preserve the codeword fidelity, the
quantity \eqref{errorrate} is a useful indicator for the
performance of a given scheme. Note that $\Lambda(F_{cw})$ is a
function of the codeword fidelity and therefore it makes sense to
use it for a comparison between different cases only for identical
values of $F_{cw}$. For our example, the fact that the coupling
constant is effectively reduced approximately $R^2/12$ times
implies that the error rate for a given value of $F_{cw}$ is also
reduced $R^2/12$ times. Similarly, the reduction of $\lambda$ by
the factor $r/6$ in the Markovian case implies a reduction of
$\Lambda$ by the same factor. We see that the effective reduction
of the error rate increases quadratically with $\kappa^2$ in the
non-Markovian case, whereas it increases only linearly with
$\kappa$ in the Markovian case.

Now let us rigorously derive the approximate solution to this model of
non-Markovian decoherence with continuous error correction.
Assuming that $\gamma \ll \kappa$ (or equivalently, $R\gg1$), the superoperator
driving the evolution of the system during a time step $\delta t$
can be written as
\begin{eqnarray}
e^{\mathcal{L}\delta t}&=&e^{\mathcal{L}_{\kappa}\delta
t}+\overset{\delta t}{\underset{0}{\int}}dt'
e^{\mathcal{L}_{\kappa}(\delta
t-t')}\mathcal{L}_{\gamma}e^{\mathcal{L}_{\kappa}t'}
+\overset{\delta t}{\underset{0}{\int}}dt'\overset{\delta
t}{\underset{t'}{\int}}dt''e^{\mathcal{L}_{\kappa}(\delta
t-t'')}\mathcal{L}_{\gamma}e^{\mathcal{L}_{\kappa}(t''-t')}\mathcal{L}_{\gamma}e^{\mathcal{L}_{\kappa}t'}+\notag\\
&+&\overset{\delta t}{\underset{0}{\int}}dt'\overset{\delta
t}{\underset{t'}{\int}}dt''\overset{\delta
t}{\underset{t''}{\int}}dt'''e^{\mathcal{L}_{\kappa}(\delta
t-t''')}\mathcal{L}_{\gamma}e^{\mathcal{L}_{\kappa}(t'''-t'')}
\mathcal{L}_{\gamma}e^{\mathcal{L}_{\kappa}(t''-t')}
\mathcal{L}_{\gamma}e^{\mathcal{L}_{\kappa}t'}+...
\label{perturbation}
\end{eqnarray}
We have denoted the Liouvillian by
$\mathcal{L}=\mathcal{L}_{\gamma}+\mathcal{L}_{\kappa}$, where
$\mathcal{L}_{\kappa}\rho=\kappa\Gamma(\rho)$, and
$\mathcal{L}_{\gamma}\rho=-i[H,\rho]$.

Let $\gamma \delta t \ll 1 \ll \kappa \delta t $. We will derive
an approximate differential equation for the evolution of
$\rho(t)$ by looking at the terms of order $\delta t$ in the
change of $\rho$ according to Eq. \eqref{perturbation}. When
$\kappa=0$, we have $d\rho/dt = \mathcal{L}_{\gamma}\rho$, so the
effect of $\mathcal{L}_{\gamma}$ on the state of the system can be
seen from Eq. \eqref{NMsystem1} with $\kappa$ taken equal to $0$.
By the action of $\exp({\mathcal{L}_{\kappa} t})$, the different
terms of the density matrix transform as follows:
$\varrho_{000,000},\varrho_{111,000},\varrho_{111,111}$ remain
unchanged, $\varrho_{100,100}\rightarrow e^{-\kappa
t}\varrho_{100,100}+(1-e^{-\kappa t})\varrho_{000,000}$,
$\varrho_{110,110}\rightarrow e^{-\kappa
t}\varrho_{110,110}+(1-e^{-\kappa t})\varrho_{111,111}$,
$\varrho_{110,001}\rightarrow e^{-\kappa
t}\varrho_{110,001}-(1-e^{-\kappa t})\varrho_{111,000}$, and all
other terms are changed as $\varrho\rightarrow e^{-\kappa t}
\varrho$. Since $\kappa \delta t \gg 1$, we will ignore terms of
order $e^{-\kappa \delta t}$. But from Eq. \eqref{perturbation} it
can be seen that all terms except
$\varrho_{000,000},\varrho_{111,000},\varrho_{000,111},\varrho_{111,111}$
will get multiplied by the factor $e^{-\kappa \delta t}$ by the
action of $\exp({\mathcal{L}_{\kappa}\delta t})$ in Eq.
\eqref{perturbation}. The integrals in Eq. \eqref{perturbation}
also yield negligible factors, since every integral either gives
rise to a factor of order $\delta t$ when the integration variable
is trivially integrated, or a factor of $1/\kappa$ when the
variable participates nontrivially in the exponent. Therefore, in
the above approximation these terms of the density matrix can be
neglected, which amounts to an effective evolution entirely within
the code space. According to Eq. \eqref{NMsystem1}, the terms
$\varrho_{000,000},\varrho_{111,000},\varrho_{111,111}$ can couple
to each other only by a triple or higher application of
$\mathcal{L}_{\gamma}$. This means that if we consider the
expansion up to the lowest nontrivial order in $\gamma$, we only
need to look at the triple integral in Eq. \eqref{perturbation}.

Let us consider the effect of $\exp({\mathcal{L}\delta t})$ on
$C_{000,000}$. Any change can come directly only from
$\varrho_{111,000}$ and $\varrho_{000,111}$. The first exponent
$e^{\mathcal{L}_{\kappa}t'}$ acts on these terms as the identity.
Under the action of the first operator $\mathcal{L}_{\gamma}$ each
of these two terms can transform to six terms that can eventually
be transformed to $\varrho_{000,000}$. They are
$\varrho_{110,000}$, $\varrho_{101,000}$, $\varrho_{011,000}$,
$\varrho_{111,100}$, $\varrho_{111,010}$, $\varrho_{111,001}$, and
$\varrho_{000,110}$, $\varrho_{000,101}$, $\varrho_{000,011}$,
$\varrho_{100,111}$, $\varrho_{010,111}$, $\varrho_{001,111}$,
with appropriate factors. The action of the second exponent is to
multiply each of these new terms by $e^{-\kappa(t''-t')}$. After
the action of the second $\mathcal{L}_{\gamma}$, the action of the
third exponent on the relevant resultant terms will be again to
multiply them by a factor $e^{-\kappa(t'''-t'')}$. Thus the second
and the third exponents yield a net factor of
$e^{-\kappa(t'''-t')}$. After the second and the third
$\mathcal{L}_{\gamma}$, the relevant terms that we get are
$\varrho_{000,000}$ and $\varrho_{100,100}$, $\varrho_{010,010}$,
$\varrho_{001,001}$, each with a corresponding factor. Finally,
the last exponent acts as the identity on $\varrho_{000,000}$ and
transforms each of the terms $\varrho_{100,100}$,
$\varrho_{010,010}$, $\varrho_{001,001}$ into
$(1-e^{-\kappa(\delta t - t''')})\varrho_{000,000}$. Counting the
number of different terms that arise at each step, and taking into
account the factors that accompany them, we obtain:
\begin{eqnarray}
C_{000,000} &\rightarrow& C_{000,000}+\overset{\delta
t}{\underset{0}{\int}}dt'\overset{\delta
t}{\underset{t'}{\int}}dt''\overset{\delta
t}{\underset{t''}{\int}}dt'''
(24e^{-\kappa(t'''-t')}-36e^{-\kappa(\delta
t-t')})C_{111,000}+\cdots\notag \\
&\approx& C_{000,000}+C_{111,000}\frac{24}{R^2}\gamma \delta
t+\textit{O}(\delta t^2).
\end{eqnarray}
Using that $C_{000,000}+C_{111,111}\approx 1$, in a similar way
one obtains
\begin{equation}
C_{111,000}\rightarrow
C_{111,000}-(2C_{000,000}-1)\frac{12}{R^2}\gamma \delta
t+\textit{O}(\delta t^2).
\end{equation}
For times much larger than $\delta t$, we can write the
approximate differential equations
\begin{gather}
\frac{d C_{000,000}}{dt}=\frac{24}{R^2}\gamma C_{111,000},\notag\\
\frac{d C_{111,000}}{dt}=-\frac{12}{R^2}\gamma
(2C_{000,000}-1).\label{approxeqn}
\end{gather}
Comparing with Eq. \eqref{sqe}, we see that the encoded qubit
undergoes approximately the same type of evolution as that of a
single qubit without error correction, but the coupling constant
is effectively decreased $R^2/12$ times. The solution of Eq.
\eqref{approxeqn} yields for the codeword fidelity
\begin{equation}
C_{000,000}(t)=\frac{1+\cos (\frac{24}{R^2}\gamma t)}{2}
\label{firstapproxsoln}.
\end{equation}
This solution is valid only with precision $\textit{O}(1/R)$ for
times $\gamma t \ll R^3$. This is because we ignored terms whose
magnitudes are always of order $\textit{O}(1/R)$ and ignored
changes of order $\textit{O}(\gamma\delta t/R^3)$ per time step
$\delta t$ in the other terms. The latter changes could accumulate
with time and become of the order of unity for times $\gamma
t\approx R^3$, which is why the approximate solution is invalid
for such times. In fact, if one carries out the expansion
\eqref{perturbation} to fourth order in $\gamma$, one obtains the
approximate equations
\begin{gather}
\frac{d C_{000,000}}{dt}=\frac{24}{R^2}\gamma C_{111,000}-\frac{72}{R^3}\gamma (2 C_{000,000}-1),\notag\\
\frac{d C_{111,000}}{dt}=-\frac{12}{R^2}\gamma (2C_{000,000}-1)
-\frac{144}{R^3}\gamma C_{111,000},\label{approxeqn2}
\end{gather}
which yield for the fidelity
\begin{equation}
C_{000,000}(t)=\frac{1+e^{-144\gamma t/R^3}\cos
(24\gamma t/R^2)}{2}.
\end{equation}
We see that in addition to the effective error process which is of
the same type as that of a single qubit, there is an extra
Markovian bit-flip process with rate $72\gamma/R^3$. This
Markovian behavior is due to the Markovian character of our
error-correcting procedure which, at this level of approximation,
is responsible for the direct transfer of weight between
$\varrho_{000,000}$ and $\varrho_{111,111}$, and between
$\varrho_{111,000}$ and $\varrho_{000,111}$. The exponential
factor explicitly reveals the range of applicability of solution
\eqref{firstapproxsoln}: with precision $\textit{O}(1/R)$, it is
valid only for times $\gamma t$ of up to order $R^2$. For times of
the order of $R^3$, the decay becomes significant and cannot be
neglected. The exponential factor may also play an important role
for short times of up to order $R$, where its contribution is
bigger than that of the cosine. But in the latter regime the
difference between the cosine and the exponent is of order
$\textit{O}(1/R^2)$, which is negligible for the precision that we
consider.

In general, the effective evolution that one obtains in the limit
of high error-correction rate does not have to approach a form
identical to that of a single decohering qubit. The reason we
obtain such behavior here is that for this particular model the
lowest order of uncorrectable errors that transform the state
within the code space is 3, and three-qubit errors have the form
of an encoded $X$ operation. Furthermore, the symmetry of the
problem ensured an identical evolution of the three qubits in the
code. For general stabilizer codes, the errors that a single qubit
can undergo are not limited to bit flips only. Therefore,
different combinations of single-qubit errors may lead to
different types of lowest-order uncorrectable errors inside the
code space, none of which in principle has to represent an encoded
version of the single-qubit operations that compose it. In
addition, if the noise is different for the different qubits,
there is no unique single-qubit error model to compare to.
Nevertheless, we will show that with regard to the effective
decrease in the error-correction rate, general stabilizer codes
will exhibit the same qualitative performance.

\section{Relation to the Zeno regime}

The effective continuous evolution \eqref{approxeqn} was derived
under the assumption that $\gamma \delta t \ll 1 \ll \kappa \delta t $.
The first inequality implies that $\delta t$
can be considered within the Zeno time scale of the system's
evolution without error correction. On the other hand, from the
relation between $\kappa$ and $\tau_c$ in \eqref{tauc} we see that
$\tau_c\ll\delta t$. Therefore, the time for implementing a weak
error-correcting operation has to be sufficiently small so that on
the Zeno time scale the error-correction procedure can be
described approximately as a continuous Markovian process. This
suggests a way of understanding the quadratic enhancement in the
non-Markovian case based on the properties of the Zeno regime.

Let us consider again the single-qubit code from Sec. II, but this
time let the error model be any Hamiltonian-driven process. We
assume that the qubit is initially in the state $|0\rangle$, i.e.,
the state of the system including the bath has the form
$\rho(0)=|0\rangle \langle 0|\otimes\rho_B(0)$. For times smaller
than the Zeno time $\delta t_Z$, the evolution of the fidelity
without error correction can be described by Eq. \eqref{Zeno}.
Equation \eqref{Zeno} naturally defines the Zeno regime in terms
of $\alpha$ itself:
\begin{equation}
\alpha\geq \alpha_Z \equiv 1-C\delta t_Z^2.
\end{equation}
For a single time step $\Delta t \ll \delta t_Z$, the change in
the fidelity is
\begin{equation}
\alpha\rightarrow \alpha-2\sqrt{C}\sqrt{1-\alpha}\Delta
t+\textit{O}(\Delta t^2).\label{singlestepdecoh}
\end{equation}
On the other hand, the effect of error correction during a time
step $\Delta t$ is
\begin{equation}
\alpha \rightarrow \alpha+\kappa (1-\alpha)\Delta t
+\textit{O}(\Delta t^2),\label{singlestepcorr}
\end{equation}
i.e., it tends to oppose the effect of decoherence.  If both
processes happen simultaneously, the effect of decoherence will
still be of the form \eqref{singlestepdecoh}, but the coefficient
$C$ may vary with time. This is because the presence of
error-correction opposes the decrease of the fidelity and
consequently can lead to an increase in the time for which the
fidelity remains within the Zeno range. If this time is
sufficiently long, the state of the environment could change
significantly under the action of the Hamiltonian, thus giving
rise to a different value for $C$ in Eq. \eqref{singlestepdecoh}
according to Eq. \eqref{C}.

Note that the strength of the Hamiltonian puts a limit on $C$, and
therefore this constant can vary only within a certain range. The
equilibrium fidelity $\alpha_*^{\rm NM}$ that we obtained for the
error model in Sec. II, can be thought of as the point at which
the effects of error and error correction cancel out. For a
general model, where the coefficient $C$ may vary with time, this
leads to a quasi-stationary equilibrium. From Eqs.
\eqref{singlestepdecoh} and \eqref{singlestepcorr}, one obtains
the equilibrium fidelity
\begin{equation}
\alpha_*^{\rm NM}\approx 1-\frac{4C}{\kappa^2}.
\end{equation}
In agreement with what we obtained in Sec. II, the equilibrium
fidelity differs from $1$ by a quantity proportional to
$1/\kappa^2$. This quantity is generally quasi-stationary and can
vary within a limited range. If one assumes a Markovian error
model, for short times the fidelity changes linearly with time
which leads to $1-\alpha_*^{\rm M}\propto 1/\kappa$. Thus the
difference can be attributed to the existence of a Zeno regime in
the non-Markovian case.

But what happens in the case of non-trivial codes? As we saw,
there the state decays inside the code space and therefore can be
highly correlated with the environment. Can we talk about a Zeno
regime then? It turns out that the answer is positive. Assuming
that each qubit undergoes an independent error process, then up to
first order in $\Delta t$ the Hamiltonian cannot map terms in the
code space to other terms without detectable errors. (This
includes both terms in the code space and terms from the hidden
part, like $\varrho_{111,000}$ in the example of the bit-flip
code.) It can only transform terms from the code space into
traceless terms from the hidden part which correspond to
single-qubit errors (like $\varrho_{100,000}$ in the same
example). Let $|\bar{0}\rangle$, $|\bar{1}\rangle$ be the two
logical codewords and $|\psi_i \rangle$ be an orthonormal basis
that spans the space of all single-qubit errors. Then in the basis
$|\bar{0}\rangle$, $|\bar{1}\rangle$, $|\psi_i \rangle$, all the
terms that can be coupled directly to terms inside the code space
are $|\bar{0}\rangle \langle \psi_i|$, $|\psi_i\rangle \langle
\bar{0}|$, $|\bar{1}\rangle \langle \psi_i|$, $|\psi_i\rangle
\langle \bar{1}|$. From the condition of positivity of the density
matrix, one can show that the coefficients in front of these terms
are at most $\sqrt{\alpha(1-\alpha)}$ in magnitude, where $\alpha$
is the code-space fidelity. This implies that for small enough
$1-\alpha$, the change in the code-space fidelity is of the type
\eqref{singlestepdecoh}, which is Zeno-like behavior. Then using
only the properties of the Zeno behavior as we did above, we can
conclude that the weight outside the code space will be kept at a
quasi-stationary value of order $1/\kappa^2$. Since uncorrectable
errors enter the code space through the action of the
error-correction procedure, which misinterprets some multi-qubit
errors in the error space, the effective error rate will be
limited by a factor proportional to the weight in the error space.
That is, this will lead to an effective decrease of the error rate
at least by a factor proportional to $1/\kappa^2$.

The accumulation of uncorrectable errors in the Markovian case is
similar, except that in this case there is a direct transfer of
errors between the code space and the visible part of the error
space. In both cases, the error rate is effectively reduced by a
factor which is roughly proportional to the inverse of the weight
in the error space, and therefore the difference in the
performance comes from the difference in this weight. The
quasi-stationary equilibrium value of the code-space fidelity
establishes a quasi-stationary flow between the code space and the
error space. One can think that this flow effectively takes
non-erroneous weight from the code space, transports it through
the error space where it accumulates uncorrectable errors, and
brings it back into the code space. Thus by minimizing the weight
outside the code space, error correction creates a ``bottleneck''
which reduces the rate at which uncorrectable errors accumulate.

Finally, a brief remark about the resources needed for quadratic
reduction of the error rate. As pointed out above, two conditions
are involved:  one concerns the rate of error correction; the
other concerns the time resolution of the weak error-correcting
operations. Both of these quantities must be sufficiently large.
There is, however, an interplay between the two, which involves
the strength of the interaction required to implement the weak
error-correcting map \eqref{wm}. Let us imagine that the weak map
is implemented by making the system interact weakly with an
ancilla in a given state, after which the ancilla is discarded.
The error-correction procedure consists of a sequence of such
interactions, and can be thought of as a cooling process which
takes away the entropy accumulated in the system as a result of
correctable errors. If the time for which a single ancilla
interacts with the system is $\tau_c$, one can verify that the
parameter $\varepsilon$ in Eq. \eqref{wm} would be proportional to
$g^2\tau_c^2$, where $g$ is the coupling strength between the
system and the ancilla. From Eq. \eqref{tauc} we then obtain that
\begin{equation}
\kappa \propto g^2\tau_c.
\end{equation}
The two parameters that can be controlled are the interaction time
and the interaction strength, and they determine the
error-correction rate. Thus if $g$ is kept constant, a decrease in
the interaction time $\tau_c$ leads to a proportional decrease in
$\kappa$, which may be undesirable. In order to achieve a good
working regime, one may need to adjust both $\tau_c$ and $g$. But
it has to be pointed out that in some situations decreasing
$\tau_c$ alone can prove advantageous, if it leads to a time
resolution revealing the non-Markovian character of an error model
which was previously described as Markovian. The quadratic
enhancement of the performance as a function of $\kappa$ may
compensate the decrease in $\kappa$, thus leading to a seemingly
paradoxical result:  better performance with a lower
error-correction rate.

\section{Conclusion}

In this paper we studied the performance of a particular
continuous quantum error-correction scheme for
non-Markovian errors. We analyzed the evolution of the
single-qubit code and the three-qubit bit-flip code in the
presence of continuous error correction for a simple non-Markovian
bit-flip error model. This enabled us to understand the workings
of the error-correction scheme, and the mechanism whereby uncorrectable errors
accumulate. The fidelity of the state with the code space in
both examples quickly reaches an equilibrium value, which can be
made arbitrarily close to $1$ by a sufficiently high rate of
error correction. The
weight of the density matrix outside the code space scales as
$1/\kappa$ in the Markovian case, while it scales as
$1/\kappa^2$ in the non-Markovian case. Correspondingly,
the rate at which uncorrectable errors accumulate
in the three-qubit code is proportional to $1/\kappa$ in the
Markovian case, and to $1/\kappa^2$ in the non-Markovian case.
These differences have the same cause, since the equilibrium
weight in the error space is closely related to the rate of
uncorrectable error accumulation.

The quadratic difference in the error weight between the Markovian
and non-Markovian cases can be attributed to the existence of a
Zeno regime in the non-Markovian case. Regardless of the
correlations between the density matrix inside the code space and
the environment, if the lowest-order errors are correctable by the
code, there exists a Zeno regime in the evolution of the
code-space fidelity. The effective reduction of the error rate
with the rate of error correction for non-Markovian error models
depends crucially on the assumption that the time resolution of
the continuous error correction is much shorter than the Zeno time
scale of the evolution {\it without} error correction. This
suggests that decreasing the time for a single (infinitesimal)
error-correcting operation can lead to an increase in the
performance of the scheme, even if the average error-correction
rate goes down.

While in this paper we have only considered codes for
the correction of single-qubit errors, our results can be
extended to other types of codes and errors as well. As long as
the error process only produces errors correctable by
the code to lowest order, an argument analogous to the one given here shows
that a Zeno regime will exist, which leads to an enhancement in the
error-correction performance. Unfortunately, it is very difficult to describe
the evolution of a system with a continuous correction protocol,
based on a general error-correction code and subject to general
non-Markovian interactions with the environment.
This is especially true if one must
include the evolution of a complicated environment in the
description, as would be necessary in general.
A more practical step in this direction might be to find
an effective description for the evolution of the reduced density
matrix of the system subject to decoherence plus error correction,
using projection techniques like the Nakajima-Zwanzig or the TCL
master equations. Since one is usually interested in the evolution
during initial times before the codeword fidelity decreases
significantly, a perturbation approach could be useful.  This is a
subject for further research.

\section*{Acknowledgements}

The authors would like to thank Kurt Jacobs for useful
information, Daniel Lidar for inspiring conversations, and Shesha
Raghunathan for his careful reading of the manuscript. This
research was supported in part by NSF Grant No. EMT-0524822.

\end{document}